\documentclass[aps,twocolumn]{revtex4-2}
\usepackage{graphicx}
\usepackage{hyperref}
\usepackage{amsmath}

\usepackage{xcolor}

\newcommand{\superref}[2]{\hyperref[#2]{#1~\ref*{#2}}}%

\newcommand{\textsupersubscript}[2]{\rlap{\textsuperscript{#1}}\textsubscript{#2}}%
\newcommand{\textsqrt}[1]{$\sqrt{\text{#1}}$}%
\newcommand{\sqrtSNN}{\textsqrt{s\textsubscript{NN}}}%
\newcommand{\MeVcc}{MeV/c\textsuperscript{2}}%
\newcommand{\AgAg}[1][1.58]{Ag(#1A~GeV)+Ag}%

\newcommand{\PiM}{$\pi$\textsuperscript{--}}%
\newcommand{\PiN}{$\pi$\textsuperscript{0}}%
\newcommand{\LambdaN}{$\Lambda$}
\newcommand{\HeliumThree}{\textsuperscript{3}He}%
\newcommand{\HeliumFour}{\textsuperscript{4}He}%
\newcommand{\Hypertriton}{\textsupersubscript{3}{\LambdaN}H}%
\newcommand{\HyperhydrogenFour}{\textsupersubscript{4}{\LambdaN}H}%
\newcommand{\HyperheliumFour}{\textsupersubscript{4}{\LambdaN}He}%

\begin{document}
    \title{Formation and lifetime measurements of light hypernuclei in Ag+Ag collisions at \sqrtSNN~=~2.55~GeV}
    
    \author{R.~Abou~Yassine$^{7,14}$, J.~Adamczewski-Musch$^{6}$, C.~Asal$^{9}$, M.~Becker$^{12}$,
A.~Belounnas$^{14}$, A.~Blanco$^{2}$, C.~Blume$^{9,6,e}$, L.~Chlad$^{15,g}$, P.~Chudoba$^{g}$,
I.~Ciepa{\l}$^{4}$
% ...
    }
    
    \affiliation{\vspace{\baselineskip}
        (HADES collaboration)\\
        \\
        \mbox{$^{1}$Ruhr-Universit\"{a}t Bochum, 44801~Bochum, Germany}\\
\mbox{$^{2}$LIP-Laborat\'{o}rio de Instrumenta\c{c}\~{a}o e F\'{\i}sica Experimental de Part\'{\i}culas, 3004-516~Coimbra, Portugal}\\
\mbox{$^{3}$AGH University of Krakow, Faculty of Physics and Applied Computer Science, 30-059~Krakow, Poland}\\
\mbox{$^{4}$Institute of Nuclear Physics, Polish Academy of Sciences, 31342~Krak\'{o}w, Poland}\\
\mbox{$^{5}$Smoluchowski Institute of Physics, Jagiellonian University of Cracow, 30-059~Krak\'{o}w, Poland}\\
\mbox{$^{6}$GSI Helmholtzzentrum f\"{u}r Schwerionenforschung GmbH, 64291~Darmstadt, Germany}\\
\mbox{$^{7}$Institut f\"{u}r Kernphysik, Technische Universit\"{a}t Darmstadt, 64289~Darmstadt, Germany}\\
\mbox{$^{8}$Institut f\"{u}r Strahlenphysik, Helmholtz-Zentrum Dresden-Rossendorf, 01314~Dresden, Germany}\\
\mbox{$^{9}$Institut f\"{u}r Kernphysik, Goethe-Universit\"{a}t, 60438 ~Frankfurt, Germany}\\
\mbox{$^{10}$Excellence Cluster 'Origin and Structure of the Universe' , 85748~Garching, Germany}\\
\mbox{$^{11}$Physik Department E62, Technische Universit\"{a}t M\"{u}nchen, 85748~Garching, Germany}\\
\mbox{$^{12}$II.Physikalisches Institut, Justus Liebig Universit\"{a}t Giessen, 35392~Giessen, Germany}\\
\mbox{$^{13}$Faculty of Science, Palack\'{y} University Olomouc, 779 00~Olomouc, Czech Republic}\\
\mbox{$^{14}$Laboratoire de Physique des 2 infinis Irene Joliot-Curie, Universite Paris-Saclay, CNRS-IN2P3, F-91405~Orsay, France}\\
\mbox{$^{15}$Nuclear Physics Institute, The Czech Academy of Sciences, 25068~Rez, Czech Republic}\\
\mbox{$^{16}$Uniwersytet Warszawski, Instytut Fizyki Do\'{s}wiadczalnej, 02-093~Warszawa, Poland}\\
\mbox{$^{17}$Warsaw University of Technology, Faculty of Physics, 00-662~Warsaw, Poland}\\
\mbox{$^{18}$Bergische Universit\"{a}t Wuppertal, 42119~Wuppertal, Germany}\\
\\
\mbox{$^{a}$ also at Instituto Politecnico de Coimbra, Instituto Superior de Engenharia de Coimbra, 3030-199~Coimbra, Portugal}\\
\mbox{$^{b}$ also at Doctoral School of Exact and Natural Sciences, Jagiellonian University, ~Cracow, Poland}\\
\mbox{$^{c}$ also at Helmholtz Research Academy Hesse for FAIR (HFHF), Campus Darmstadt, 64390~Darmstadt, Germany}\\
\mbox{$^{d}$ also at Technische Universit\"{a}t Dresden, 01062~Dresden, Germany}\\
\mbox{$^{e}$ also at Helmholtz Research Academy Hesse for FAIR (HFHF), Campus Frankfurt, 60438~Frankfurt am Main, Germany}\\
\mbox{$^{f}$ also at Charles University, Faculty of Mathematics and Physics, 12116~Prague, Czech Republic}\\
\mbox{$^{g}$ also at Czech Technical University in Prague, 16000~Prague, Czech Republic}\\
\mbox{$^{h}$ also at University of Wroc{\l}aw, 50-204 ~Wroc{\l}aw, Poland}        
    }


    \begin{abstract}
       We present the first observation of \Hypertriton\ and \HyperhydrogenFour\ in Ag+Ag collisions at \sqrtSNN~=~2.55~GeV, emitted around mid-rapidity. The hypernuclei are reconstructed via their two-body decay channels and identified through their weak-decay topology, employing an artificial neural network for enhanced discrimination. 
       The analysis methodology is validated using \LambdaN~hyperons. The resulting rapidity distributions, dN/dy, exhibit a bell shape centered at mid-rapidity. The yield of \HyperhydrogenFour\ is equal to or exceeds that of \Hypertriton, which contrasts the measurement from the STAR collaboration at \sqrtSNN~=~3~GeV and is consistent with a scenario in which hypernuclei receive feed-down from excited states.
       The data enable a high-precision measurement of the hypernuclei lifetimes. 
       For the \Hypertriton, a lifetime of \(\tau_{{}^{3}_{\Lambda}\mathrm{H}} = (239 \pm 23{\mathrm{(stat)}} \pm 18{\mathrm{(sys)}})\,\mathrm{ps}\), is extracted, consistent on the 1$\sigma$ level with that of the free \LambdaN. In contrast, the \HyperhydrogenFour\ lifetime of \(\tau_{{}^{4}_{\Lambda}\mathrm{H}} = (209 \pm 7{\mathrm{(stat)}} \pm 10{\mathrm{(sys)}})\,\mathrm{ps}\), shows a 4.5 $\sigma$ deviation from the free \LambdaN\ lifetime. The results consolidate the available world data.
    \end{abstract}

    \maketitle

\section{Introduction}

Hypernuclei extend the nuclear chart into the strange sector and thus offer unique opportunities to study the hyperon–nucleon (Y–N) interaction.  
This interaction plays a crucial role in the equation of state (EOS) of nuclear matter at high baryon densities~\cite{Pandharipande:1971up,Bethe:1974gy,Glendenning:1991es,Balberg:1997yw,bib:Hypernuclei2,bib:Hypernuclei1}.  
For example, the appearance of hyperons in neutron stars is expected to soften the EOS, thereby affecting the range of stable mass–radius configurations~\cite{Lattimer:2004pg}, and may conflict with the observation of $\sim$2\,$M_\odot$ neutron stars.

Since the lifetimes of hypernuclei are sensitive to the Y–N interaction~\cite{Dalitz:1962eb,Kamada:1997rv}, measurements of their decay curves provide experimental access to study this interaction under various conditions. Heavy-ion collisions enable the formation of hypernuclei in environments of finite temperature and baryon density above nuclear ground state conditions, offering a setting for studying the Y--N interaction.

The lightest known hypernucleus, the \Hypertriton, has a mass of  2991~\MeVcc\ and a nuclear binding energy of only about 0.8~MeV per baryon~\cite{bib:Hypernuclei2}.  
Because the \Hypertriton\ has an exceptionally small $\Lambda$ separation energy, it is effectively a weakly bound state with a large spatial extension \cite{Braun-Munzinger:2018hat}. In this dilute configuration the $\Lambda$ samples only low effective nucleon density. Consequently, the \Hypertriton\ lifetime is expected to be close to the free $\Lambda$ value.

However, multiple experimental measurements of the \Hypertriton\ lifetime have yielded values significantly shorter than the free \LambdaN\ lifetime~\cite{bib:STAR_Hypertriton2010,bib:HypHI_Hypertriton2013,bib:ALICE_Hypertriton2016,bib:STAR_Hypertriton2018}, giving rise to the so-called \Hypertriton~lifetime puzzle. More recent results, in contrast, report lifetimes that are only slightly below or even consistent with that of the free \LambdaN\ within uncertainties~\cite{bib:ALICE_Hypertriton2019,bib:STAR_Hypertriton2022,ALICE:2022rib}.

The \Hypertriton\ has six mesonic decay channels, which result from a combination of the two primary decay modes of the \LambdaN\ hyperon with the possibility of no, partial, or full nuclear breakup in the final state. The most probable decay modes involve a \textit{partial nuclear breakup}:
\Hypertriton~$\to$~d~+~p~+~\PiM\ (BR~$\approx$~40\%)
and \Hypertriton~$\to$~d~+~n~+~\PiN\ (BR~$\approx$~20\%).
Decays with \textit{no nuclear breakup} proceed via
\Hypertriton~$\to$~\HeliumThree~+~\PiM\ (BR~$\approx$~25\%) and
\Hypertriton~$\to$~t~+~\PiN\ (BR~$\approx$~13\%).
Channels involving \textit{complete nuclear breakup}, \Hypertriton~$\to$~p~+~p~+~n~+~\PiM\ and \Hypertriton~$\to$~p~+~n~+~n~+~\PiN, as well as nonmesonic decays, have only negligible branching ratios. A detailed theoretical review of the \Hypertriton\ decay modes can be found in~\cite{bib:HypertritonDecay}.

The \HyperhydrogenFour\ decay channels are similar to the ones of the \Hypertriton. However, since the only stable nucleus with mass number $A=4$ is \HeliumFour, there exists only \textit{one decay channel without nuclear breakup}:
\HyperhydrogenFour~$\to$~\HeliumFour~+~\PiM\ (BR~$\approx$~50\%).
Two additional channels involve \textit{partial nuclear breakup}:
\HyperhydrogenFour~$\to$~t~+~p~+~\PiM\ (BR~$\approx$~33\%)
and \HyperhydrogenFour~$\to$~t~+~n~+~\PiN\ (BR~$\approx$~17\%)~\cite{Outa:1995es}.
All other decay modes, including those with complete nuclear breakup, have negligible branching ratios. Basic theoretical estimates predict a reduction of the \HyperhydrogenFour~lifetime by approximately 30\% compared to the free \LambdaN\ lifetime~\cite{Gal:2021ulo}.

In heavy-ion collisions at center-of-mass energies of a few GeV per nucleon pair, nucleons from the colliding nuclei experience significant stopping, causing a baryon dominated fireball. This environment provides favorable conditions for the formation and emission of light nuclei. Simultaneously, hadrons containing strange quarks are produced near their kinematic production thresholds. Consequently, their yields exhibit a steep increase with rising \sqrtSNN. The interplay between these effects results in an maximum of hypernucleus yields around \sqrtSNN~=~3–5~GeV~\cite{Andronic:2010qu} per collision.

Following first attemps by the FOPI collaboration ~\cite{ZHang} and our previously published upper production limit in Ar+KCl collisions ~\cite{bib:HADES_HypertritonArKCl}, we report the first observation of \Hypertriton\ and \HyperhydrogenFour\ emitted from central Ag+Ag collisions at \sqrtSNN~=~2.55~GeV around mid-rapidity. 

This paper is organized as follows. In Section~\ref{sec:Rec}, we introduce the experimental setup and describe the reconstruction methods. Section~\ref{sec:multidiff} presents the multi-differential yields and discusses the c.m. energy dependence at mid-rapidity. Section~\ref{sec:lifetime} is dedicated to the decay curve analysis and the extraction of lifetimes. We conclude in Section~\ref{sec:summary} with a summary of our findings.

\section{Experimental Setup and Reconstruction}\label{sec:Rec}

The data analyzed in this work were recorded with the High Acceptance Di-Electron Spectrometer (HADES) located at the GSI Helmholtz Centre for Heavy-Ion Research in Darmstadt, Germany. HADES is a charged-particle detector composed of a six-coil toroidal magnet surrounding the beam axis and six identical detection sectors placed between the coils, providing nearly complete azimuthal coverage. Each sector is equipped with a Ring-Imaging Cherenkov (RICH) detector, followed by four Multi-Wire-Drift Chambers (MDCs)—two placed upstream and two downstream of the magnetic field—along with a scintillator hodoscope (TOF) and a Resistive Plate Chamber (RPC). Downstream of the tracking system, an Electromagnetic Calorimeter (ECAL) for photon measurements and a forward hodoscope for event-plane determination are employed. A detailed description of the detector system is available in~\cite{HADES:2009aat}.

The MDCs provide charged-particle tracking, while TOF and RPC are employed for time-of-flight measurements in conjunction with a diamond start detector positioned upstream of a 15-fold segmented Ag target. The trigger condition was based on hit multiplicities observed in the TOF and RPC detectors. A total of $2.6 \times 10^9$ \AgAg\ collisions were analyzed, corresponding to the $0$--$25\%$ most central events.

Particle identification was achieved by applying loose selection criteria to the correlation between reconstructed momentum and particle velocity. For \HeliumThree\ and \HeliumFour, the specific energy loss ($dE/dx$) measured in the MDCs was used in addition to suppress background from singly charged particles. Further selection criteria were applied to track quality based on the Runge-Kutta track fitting algorithm.

The reconstruction of \Hypertriton\ and \HyperhydrogenFour\ candidates was performed by identifying “off-vertex” two-body decay topologies. Background suppression was achieved by requiring a maximal distance of closest approach (DCA) between the primary event vertex and the daughter tracks (\HeliumThree\ or \HeliumFour\ and \PiM), as well as a minimal DCA between the decay vertex and the reconstructed hypernucleus trajectory. A minimum opening angle between the decay products was also required. To further enhance the signal purity, an artificial neural network (aNN)~\cite{bib:TMVA} was employed. The network was trained in a supervised manner using simulated \Hypertriton\ and \HyperhydrogenFour\ signal samples, along with background distributions generated via the mixed-event technique. The reconstruction procedure follows the method described in~\cite{bib:SimonSpies_PhD} and has been previously validated for \LambdaN\ hyperons~\cite{bib:LamK0sAuAu}. In parallel, \LambdaN\ hyperons were reconstructed from the same data sample and serve as reference signals throughout the analysis.

\par
    \begin{figure}
        \centering
        \includegraphics[width=0.9\linewidth]{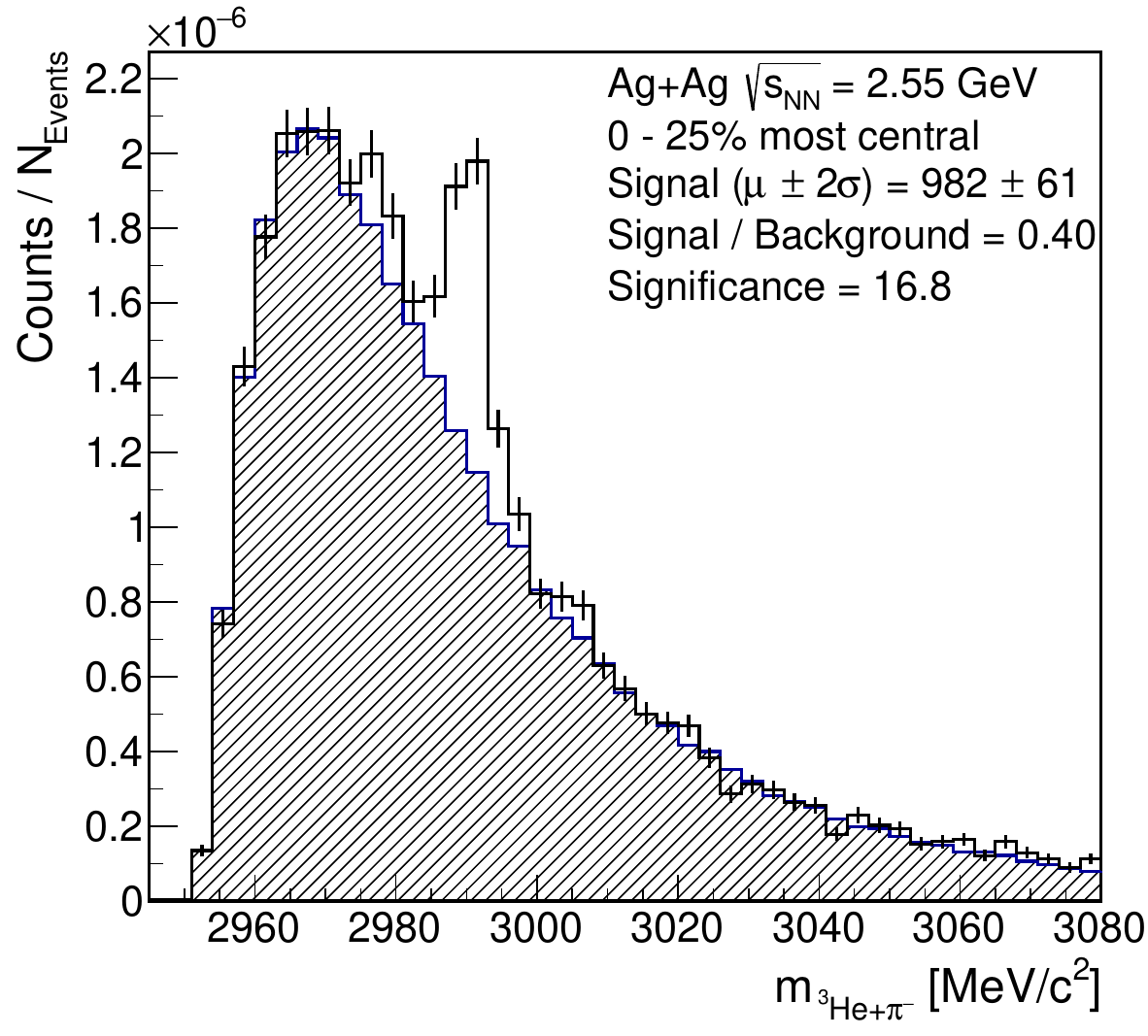}
        \includegraphics[width=0.9\linewidth]{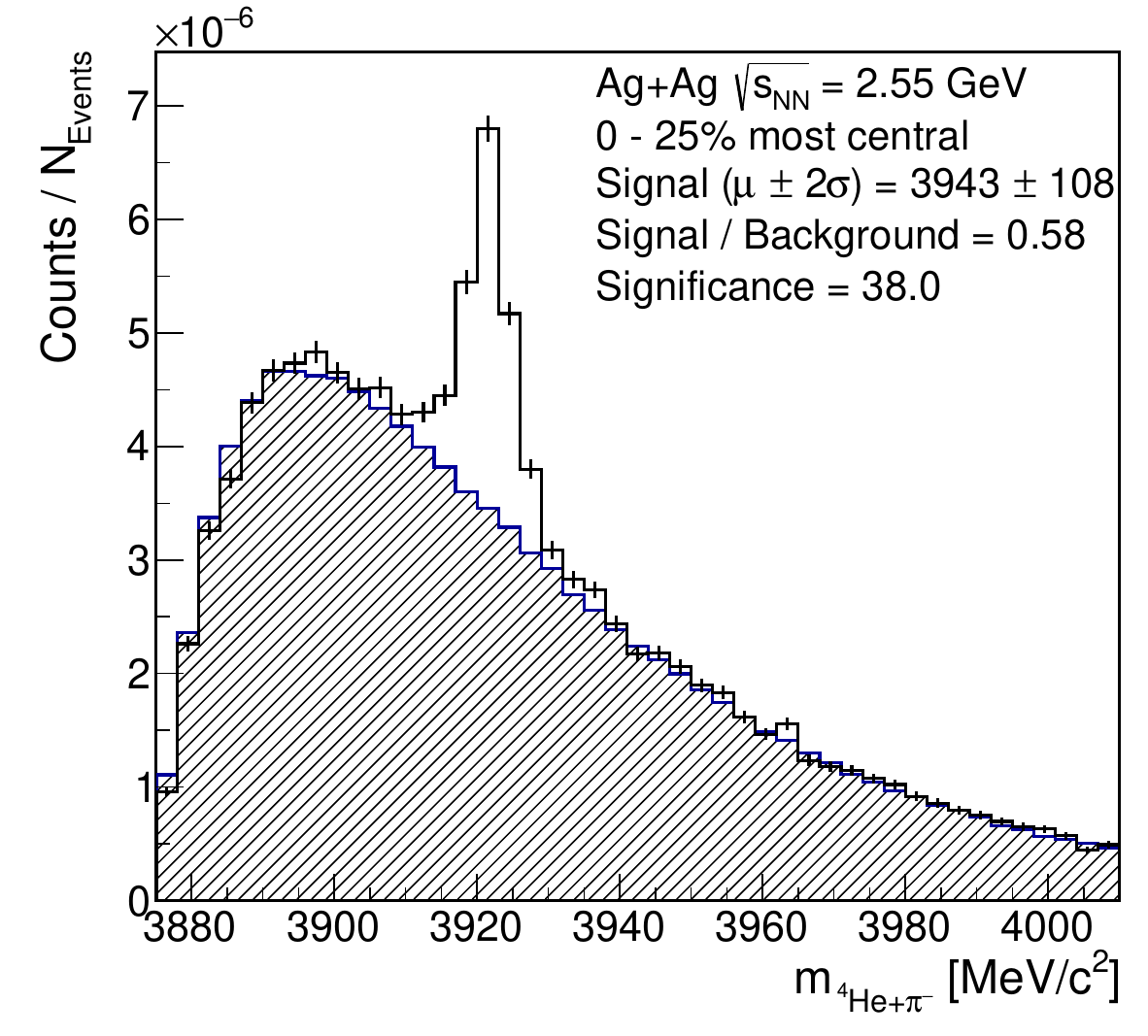}
        \caption{Upper panel: Invariant mass spectra for the same-event (black) and mixed-event (blue) combinations of \HeliumThree- and \PiM-tracks. The mixed-event is normalized to the same-event outside of a \textpm 5$\sigma$ region around the nominal \Hypertriton\ mass. Lower panel: Same, but for combinations of \HeliumFour- and \PiM-tracks.}
        \label{fig:RecSig}
    \end{figure}
 
The remaining combinatorial background in the invariant mass spectra is estimated using the mixed-event technique. The mixed-event distribution is normalized to the same-event distribution outside a $\pm 5\sigma$ window around the nominal masses.

Invariant mass distributions for same-event (black) and mixed-event (blue) combinations of \HeliumThree- and \PiM-tracks are shown in the upper panel of \superref{Fig.}{fig:RecSig}. The \Hypertriton\ signal is clearly visible with, a significance ($S/\sqrt{S+B}$) of 16.8 and a signal-to-background ratio of 0.4 within a $\pm2\sigma$ mass window. Subtracting the mixed-event background and integrating within this window results in approximately 1000 \Hypertriton\ signal candidates.

The corresponding distributions for the \HyperhydrogenFour\ candidates, obtained from \HeliumFour- and \PiM-track combinations, are shown in the lower panel of \superref{Fig.}{fig:RecSig}. The \HyperhydrogenFour\ signal exhibits a significance of 38.0 and a signal-to-background ratio of 0.58 within a $\pm2\sigma$ mass window. After background subtraction, approximately 4000 \HyperhydrogenFour\ signal candidates are extracted.

\section{Transverse Momentum Spectra, Rapidity Distributions and Yields} \label{sec:multidiff}

\begin{figure}
    \centering
    \includegraphics[width=0.88\linewidth]{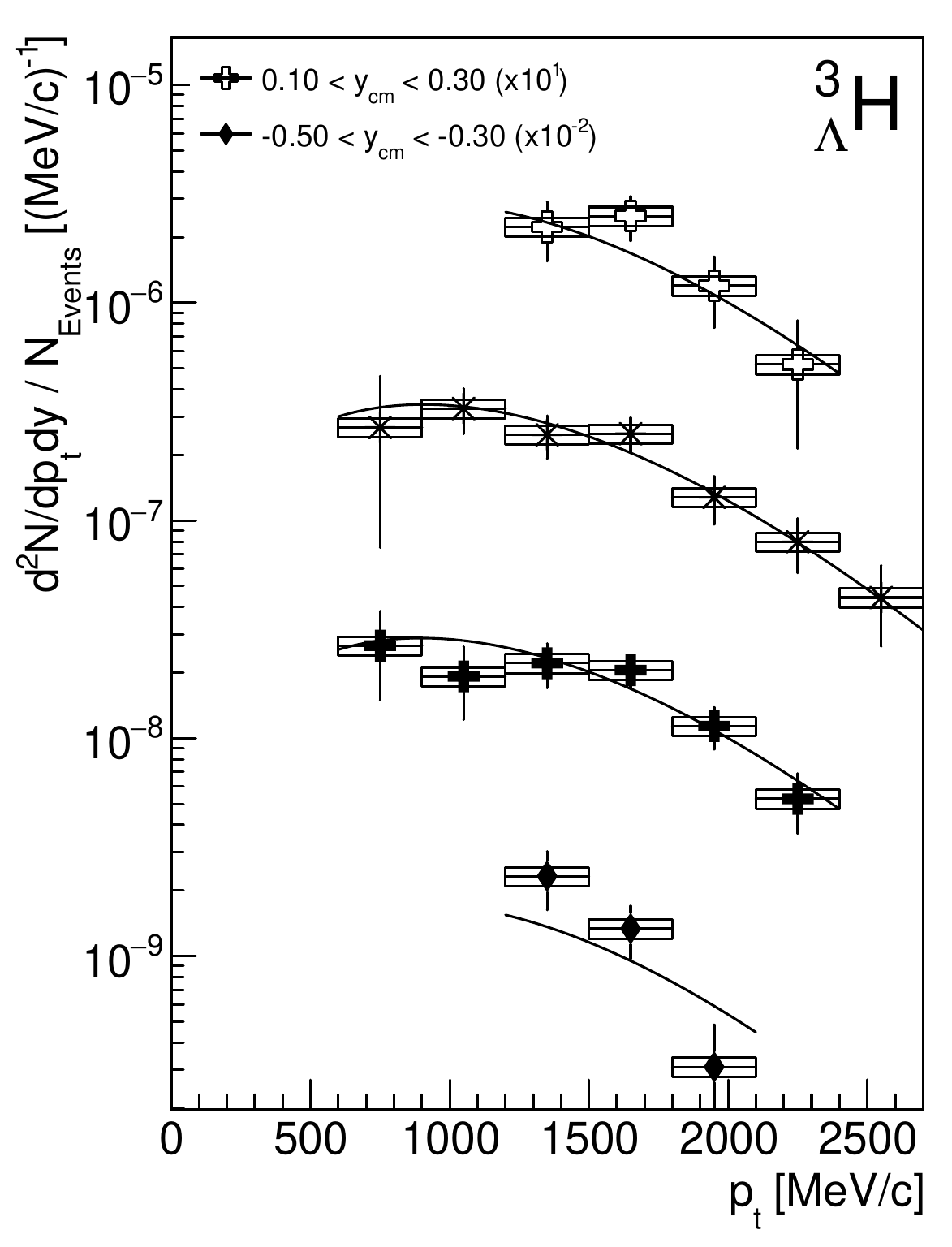}
    \includegraphics[width=0.88\linewidth]{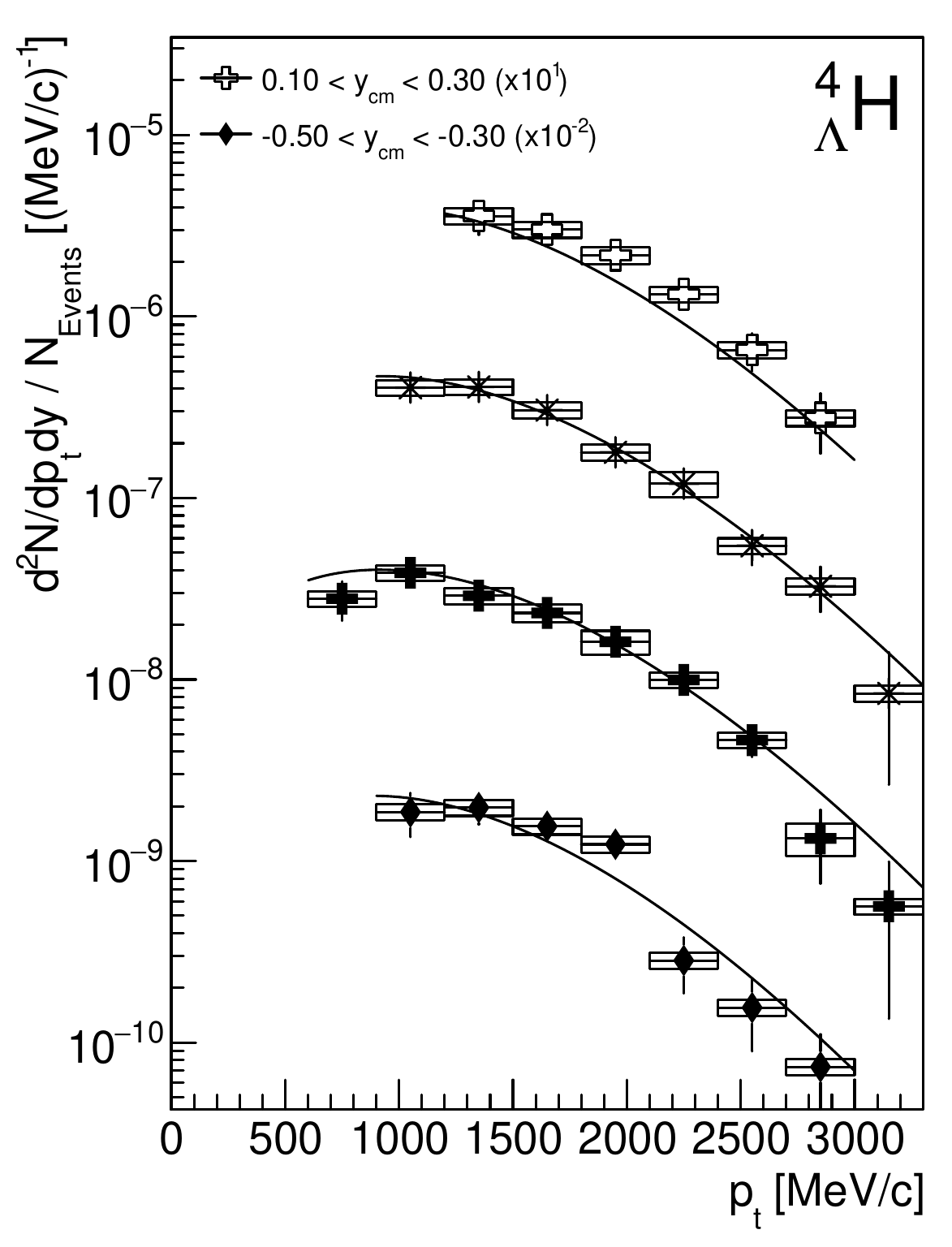}
    \caption{Acceptance- and efficiency-corrected transverse momentum $p_{\mathrm{T}}$ spectra for subsequent slices of rapidity $y_{\mathrm{cm}}$ for \Hypertriton{} (upper panel) and \HyperhydrogenFour{} (lower panel) nuclei. Displayed is the yield per event, per transverse momentum interval, and per unit of rapidity for the 0--25\% most central \AgAg{} collisions. For a better representation, the spectra are scaled by consecutive
factors of 10 for each rapidity slice. The boxes indicate the systematic uncertainties, while the curves correspond to fits using Eq.~(1), see text for details.
    }
    \label{fig:Spec}
\end{figure}
The available statistics enables for both hypernuclei a multi-differential analysis  as a function of rapidity $y$ and transverse momentum $p_T$.
\Hypertriton\ yields are determined in four rapidity intervals, covering the center-of-mass rapidity $y_{\mathrm{cm}}$ between $-0.5$ and $0.3$ in steps of $0.2$ units in rapidity, and up to seven $p_{\mathrm{T}}$-intervals ranging from 600 to 2700~MeV/$c$ in steps of 300~MeV/$c$.
\par
\indent\HyperhydrogenFour\ yields are determined analogously in four rapidity intervals, covering $y_{\mathrm{cm}}$ between $-0.5$ and $0.3$ in steps of $0.2$ units, and up to nine $p_{\mathrm{T}}$-intervals ranging from 600 to 3300~MeV/$c$, again in steps of 300~MeV/$c$.
\par
\indent
The obtained significance and signal-to-background ratios vary between 1.4--5.8 and 0.2--1.1, respectively, for the \Hypertriton{}, and between 1.3--12.1 and 0.2--0.9 for the \HyperhydrogenFour{}.
The count rates are corrected in each phase-space cell for acceptance and efficiency 
losses. These corrections are determined by embedding hypernuclei signals, simulated with Geant 3.21~\cite{bib:GEANT3} in combination with a detailed description of the detector response, into experimental data.
The simulation dynamically accounts for beam intensity fluctuations within a given spill on an event-by-event basis. Since the $\delta$-electron flux produced in the target is proportional to the instantaneous beam intensity, they increase the occupancy in the first two MDCs and thus affect the track reconstruction efficiency.
As input to the simulation, thermal-like distributions according to Eq.~(1) of \Hypertriton{} and \HyperhydrogenFour{} are used. Their parameters are iteratively adjusted to match those extracted from fits to the experimental data, which are employed for the extrapolation in transverse momentum and rapidity, as discussed below.
For the \Hypertriton{}, a lifetime equal to that of a free $\Lambda$ hyperon and a branching ratio of 25\% for the $\mathrm{{}^3He} + \pi^-$ decay channel are assumed. For the \HyperhydrogenFour{}, a lifetime equal to that of a free $\Lambda$ hyperon and a branching ratio of 50\% for the $\mathrm{{}^4He} + \pi^-$ decay channel are used. The simulated hypernuclei are subjected to the same reconstruction and analysis steps as the experimental data. 

The acceptance- and efficiency-corrected $p_{\mathrm{T}}$ distributions for successive slices of center-of-mass rapidity for \Hypertriton{} (upper panel) and \HyperhydrogenFour{} nuclei (lower panel) are presented in \superref{Fig.}{fig:Spec}. 

For the extrapolation of the multi-differential production yields \(d^{2}N/(dp_{\mathrm{T}}\,dy)\) to regions in rapidity and transverse momentum not covered by the measurement, the data are fitted using functions motivated by a thermal ansatz involving an effective temperature \(T_{\mathrm{eff}}\):

\begin{equation}
    \begin{split}
        \frac{d^{2}N}{dp_{\mathrm{T}}\,dy} =&\ C\cdot p_{\mathrm{T}} \left( E(p_{\mathrm{T}}, y-\eta) \exp\left(-\frac{E(p_{\mathrm{T}}, y-\eta)}{T_{\mathrm{eff}}}\right) \right.\\
        &+ \left. E(p_{\mathrm{T}}, y+\eta) \exp\left(-\frac{E(p_{\mathrm{T}}, y+\eta)}{T_{\mathrm{eff}}}\right) \right),
    \end{split}w
    \label{eq:thermal_fit}
\end{equation}

where \(C\) is a normalization constant, and \(E(p_{\mathrm{T}}, y)\) represents the energy of the particle as a function of transverse momentum and rapidity.
The dimensionless parameter \(\eta\) accounts for the longitudinal anisotropy arising from the incomplete stopping of the colliding nucleons in the interaction zone as well as from radial flow.  

To achieve stable fits of the yield spectra within the available statistics, only the normalization constant \(C\) is treated as a free parameter for each \(d^{2}N/(dp_{\mathrm{T}}\,dy)\) spectrum, while \(T_{\mathrm{eff}}\) is fitted as a global parameter across all phase-space cells, assuming a \(1/\cosh(y)\) dependence. 
The parameter \(\eta\) is fixed to a value of 0.15, determined from a \(\chi^{2}\) scan with \(T_{\mathrm{eff}}\) held constant.

The systematic uncertainties are estimated from variations of the \(d^{2}N/(dp_{\mathrm{T}}\,dy)\) yields under changes of the particle identification (PID) cuts, topological selection criteria, the parameter \(\eta\) of the fit function, and the interaction cross section of the hypernuclei with the detector material.
The PID selections in the \(\beta\) versus momentum and \(\mathrm{d}E/\mathrm{d}x\) versus momentum distributions are varied within \(\pm1\sigma\). 
The topological selections are modified using three additional configurations—loose, medium, and tight—excluding the aNN support.

The parameter \(\eta\) is varied within a reasonable range of \(\pm 0.05\).
Interactions of hypernuclei with detector material occur primarily in the beam pipe (aluminum) and in the RICH radiator gas (isobutane). The hypernuclear breakup cross section is varied from 0 to 4~barn for aluminum (steps: 0, 2, 4~b), from 0 to 1~barn for hydrogen (0, 0.5, 1~b), and from 0 to 3~barn for carbon (0, 1.5, 3~b). The hydrogen and carbon interaction probabilities are weighted according to the mole fractions of the isobutane gas components.
The resulting variations in yield, which exceed the statistical uncertainties, are added in quadrature to obtain the total systematic uncertainty, which amounts in total to 10\% for both hypernuclei.

The estimated statistical and systematic uncertainties are consistent with the observed deviations between forward and backward rapidity.

An additional overall normalization uncertainty of 5\% arises from the selection of the event class corresponding to the 0–25\% most active collisions. This contribution is not included in the systematic uncertainties shown in the figures.

The extracted \(T_{\mathrm{eff}}\) values are \((242~\pm~20{\mathrm{(stat)}}~\pm~19{\mathrm{(sys)}})\)~MeV for \Hypertriton{}, and \((202~\pm~6{\mathrm{(stat)}}~\pm~10{\mathrm{(sys)}})\)~MeV for \HyperhydrogenFour{}. 
The corresponding \(d^{2}N/(dp_{\mathrm{T}}\,dy)\) distributions for the \Hypertriton{} (upper panel) and the \HyperhydrogenFour{} (lower panel) are displayed as curves in \superref{Fig.}{fig:Spec}.

The \(dN/dy\) values do not include systematic uncertainties arising from the model-dependent \(p_{\mathrm{T}}\)-extrapolations. Therefore, we recommend a direct comparison of the multi-differential yields \(d^{2}N/(dp_{\mathrm{T}}\,dy)\) with phenomenological models.

\begin{figure}
        \centering
        \includegraphics[width=\linewidth]{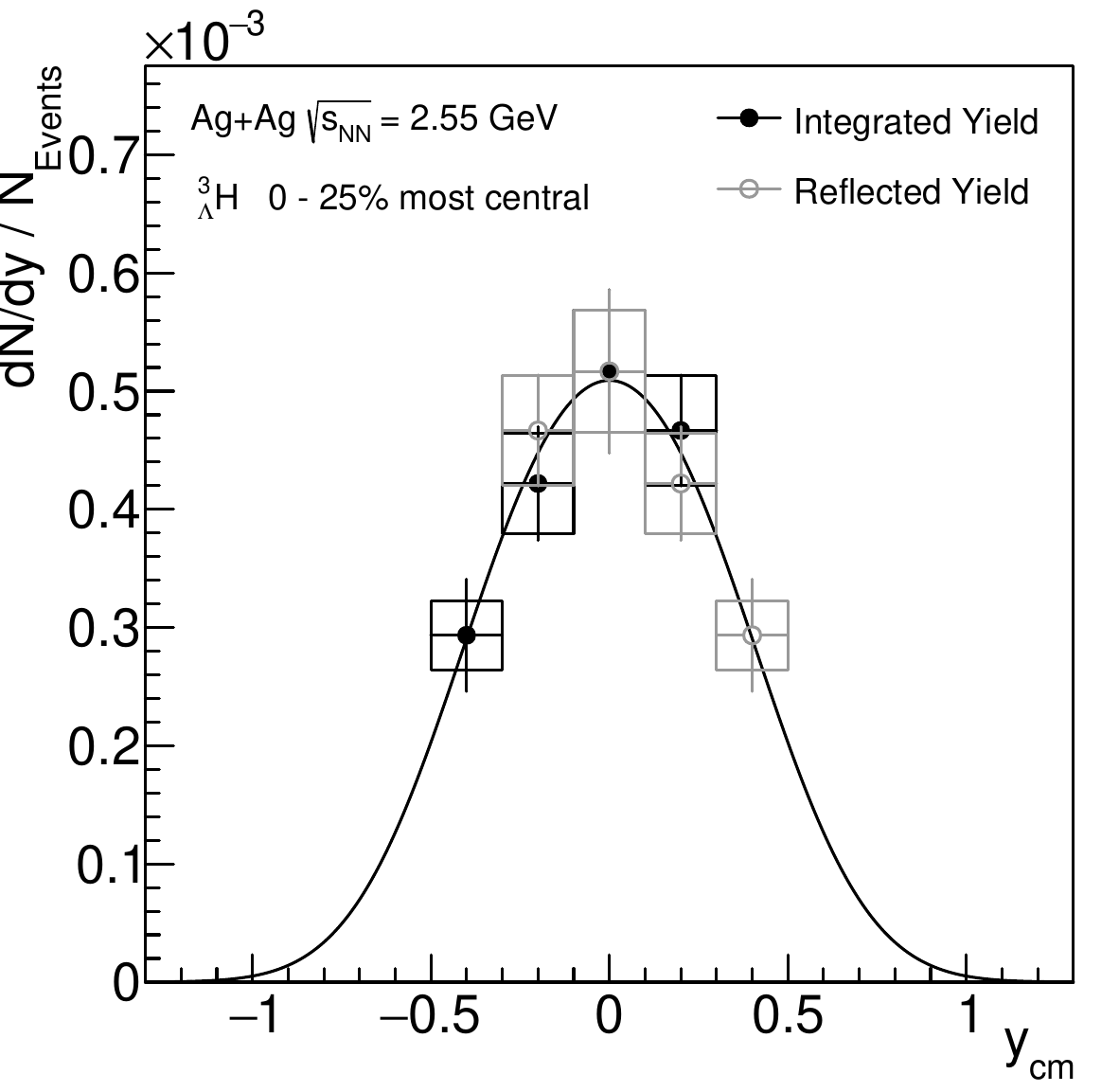}
        \includegraphics[width=\linewidth]{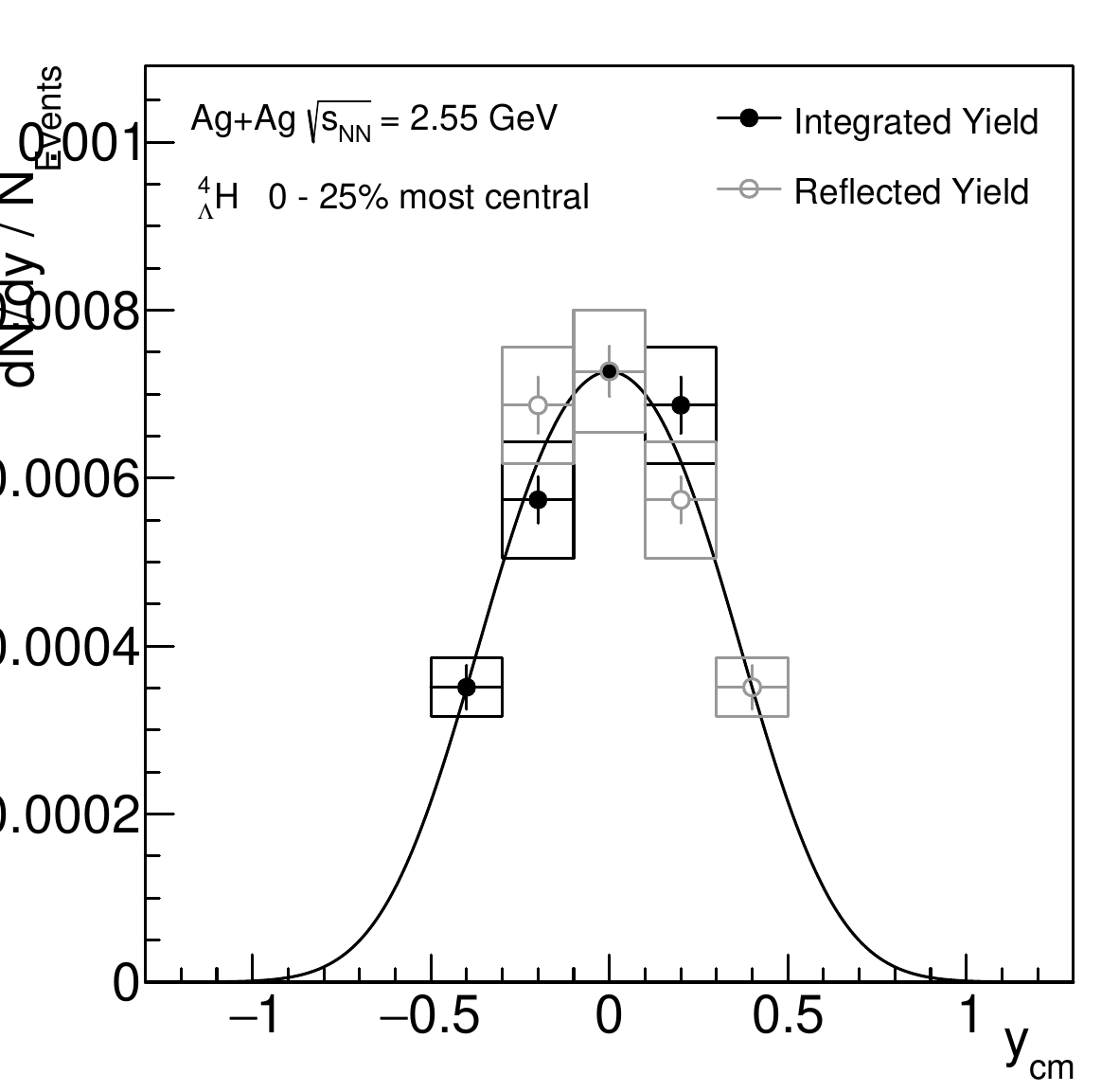}
        \caption{Rapidity distributions $dN/dy$ for the \Hypertriton{} (upper panel) and the \HyperhydrogenFour{} (lower panel), obtained by integrating the data as a function of transverse momentum $p_{\mathrm{T}}$ and using fit functions to extrapolate into $p_{\mathrm{T}}$ regions not covered by our detector. The boxes indicate the systematic uncertainties, while the curves correspond to distributions motivated by the thermal ansatz fitted to the data; see text for details.}
\label{fig:dNdy}
\end{figure}
The rapidity distributions \(dN/dy\) for the \Hypertriton{} (upper panel) and the \HyperhydrogenFour{} (lower panel), displayed in \superref{Fig.}{fig:dNdy}, are obtained by integrating the data as a function of transverse momentum, extrapolated using the presented fit functions into the \(p_{\mathrm{T}}\) regions not covered by our detector.
Both hypernuclei exhibit bell-shaped \(dN/dy\) distributions. 
The total production rates are obtained by integrating the \(dN/dy\) distributions as a function of center-of-mass rapidity, supplemented by fit functions for extrapolation \(y_{\mathrm{cm}}\) regions not covered by our detector.
For this purpose, we adopt the thermal ansatz introduced in Eq.~(1). In the case of a purely thermal source, the rapidity distribution is given by:

\begin{equation}
    \begin{split}
        \frac{dN_{\mathrm{th}}}{dy} &= C \cdot T_{\mathrm{eff}} \cdot \exp\left(-\frac{m_0 \cosh(y)}{T_{\mathrm{eff}}} \right) \\
        &\quad \cdot \left( m_0^2 + 2 m_0 \frac{T_{\mathrm{eff}}}{\cosh(y)} + 2 \frac{T_{\mathrm{eff}}^2}{\cosh^2(y)} \right),
    \end{split}
\end{equation}
As before, the parameter \(\eta\), accounting for the longitudinal anisotropy caused by incomplete stopping of the colliding nucleons and collective longitudinal flow, is introduced, yielding the final function used for extrapolation:

\begin{equation}
    \frac{dN}{dy} = \left.\frac{dN_{\mathrm{th}}}{dy}\right|_{y - \eta} + \left.\frac{dN_{\mathrm{th}}}{dy}\right|_{y + \eta}
\end{equation}
The resulting production rates are \((4.8~\pm~0.5{\mathrm{(stat)}}~\pm~0.5{\mathrm{(sys)}}~\pm~0.2{\mathrm{(norm)}}) \times 10^{-4}\) \Hypertriton{} per event and \((6.1~\pm~0.2{\mathrm{(stat)}}~\pm~0.6{\mathrm{(sys)}}~\pm~0.3{\mathrm{(norm)}}) \times 10^{-4}\) \HyperhydrogenFour{} per Ag+Ag collision within the above defined centrality class.

The yield of \HyperhydrogenFour\ is equal to or exceeds that of \Hypertriton, which contrasts the measurement from the STAR collaboration at \sqrtSNN~=~3~GeV.

This observation might be related to the existence of an excited state of hypernuclei with mass number \(A=4\)~\cite{Bedjidian:1976zh,CERN-Lyon-Warsaw:1979ifx}, which decays via photon emission to the ground state and is not observed in the case of the \Hypertriton{}.

\par
In a statistical-thermal approach to particle emission, the multiplicity \(M_{b_i}\) (without feed-down) of baryons of species \(i\) is given by:

\begin{equation}
    M_{b_i} = g_i V \int \frac{d^3p}{(2\pi)^3} \exp\left( -\frac{E_i - B \mu_B}{k_B T} \right) \times F_{S_i},
    \label{2_eq}
\end{equation}
where \(T\) is the temperature, \(V\) the volume, \(\mu_B\) the baryochemical potential, \(g_i = 2J_i + 1\) the degeneracy factor, \(B\) the baryon number, \(E_i = \sqrt{m_i^2 + p^2}\) the energy of the corresponding baryon, and \(F_{S_i}\) a fugacity factor accounting for additional conservation laws.

Neglecting the mass difference between \HyperhydrogenFour{} and its excited state \HyperhydrogenFour*, and using \(g = 2J + 1\), a ratio of excited- to ground-state yields of 3:1 is found~\cite{BenD_SQM22}.
Therefore, at high values of the baryochemical potential (or equivalently at low \(\sqrt{s_{\mathrm{NN}}}\)), where the difference \(E_i - \mu_B\) becomes small and no longer compensates for the additional contributions from excited states in the case of the \HyperhydrogenFour{}, similar or even higher yields of \HyperhydrogenFour{} compared to the \Hypertriton{} are expected.

\begin{figure}
        \centering
        \includegraphics[width=\linewidth]{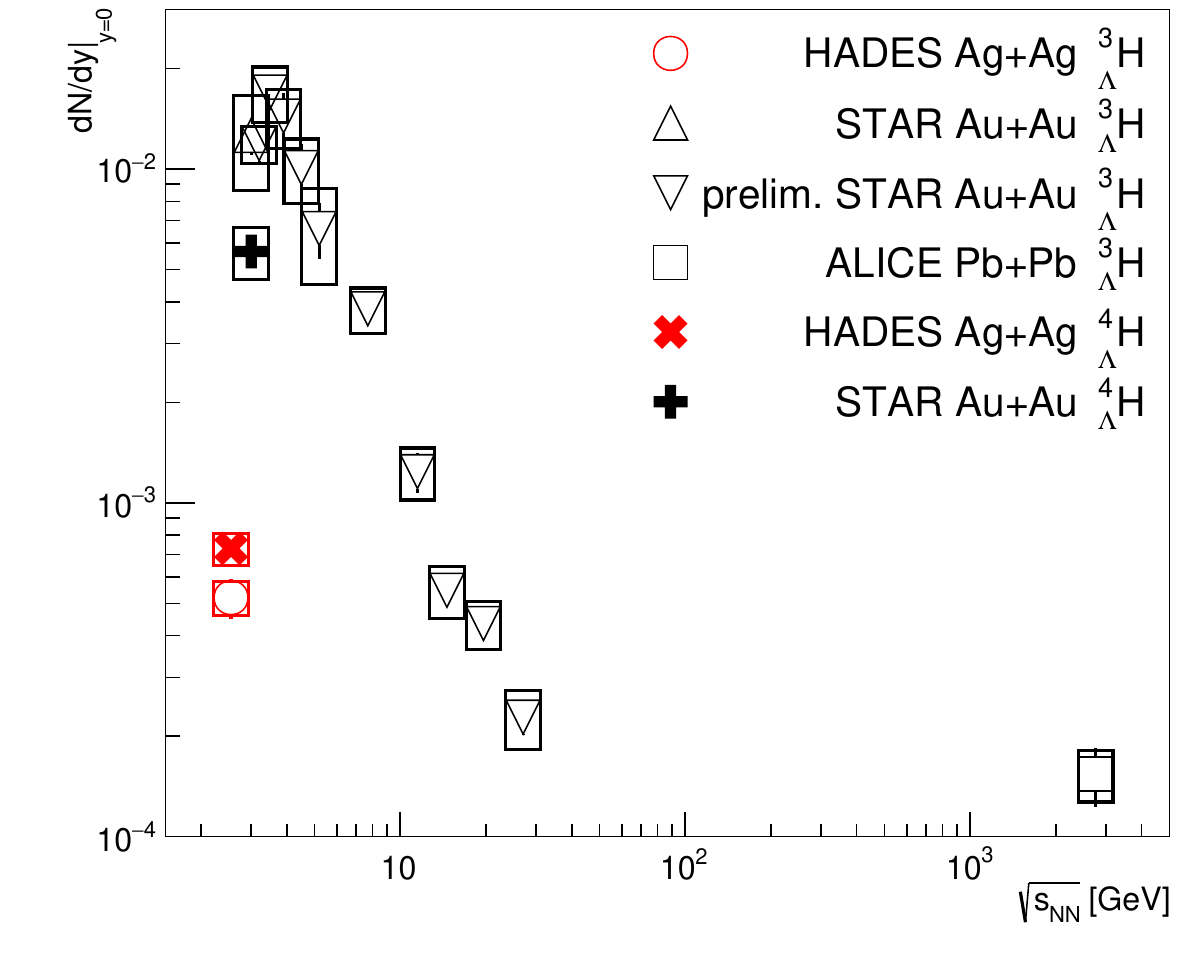}
        \caption{Compilation of energy excitation function of \Hypertriton{} (empty symbols) and \HyperhydrogenFour{} (fat symbols) yields emitted around mid-rapidity in central heavy-ion collisions from \cite{bib:STAR_Hypertriton2022,bib:STAR_Hypertriton2025,bib:ALICE_Hypertriton2019}, the preliminary STAR data \cite{bib:STAR_Hypertriton2025}, and the present work.}
        
        \label{fig:ex}
\end{figure}
   
World data on \Hypertriton{} yields emitted around midrapidity in central heavy-ion collisions~\cite{bib:STAR_Hypertriton2022,bib:ALICE_Hypertriton2019,bib:STAR_Hypertriton2025}, are shown in \superref{Fig.}{fig:ex} as a function of \(\sqrt{s_{\mathrm{NN}}}\). 
The data exhibit a clear maximum at \(\sqrt{s_{\mathrm{NN}}} \approx 3\) GeV, consistent with the expectation of an intermediate maximum in the hypernuclear excitation function.

In contrast, the energy dependence of \HyperhydrogenFour{} yields remains poorly constrained due to the limited available data. To date, only one additional data point at \(\sqrt{s_{\mathrm{NN}}} \approx 3\) GeV has been reported~\cite{bib:STAR_Hypertriton2022}, as shown in \superref{Fig.}{fig:ex}, leaving significant room for future measurements, for example at the upcoming CBM experiment at GSI/FAIR.

\section{Decay Curves} \label{sec:lifetime}
    \begin{figure}
        \centering
        \includegraphics[width=\linewidth]{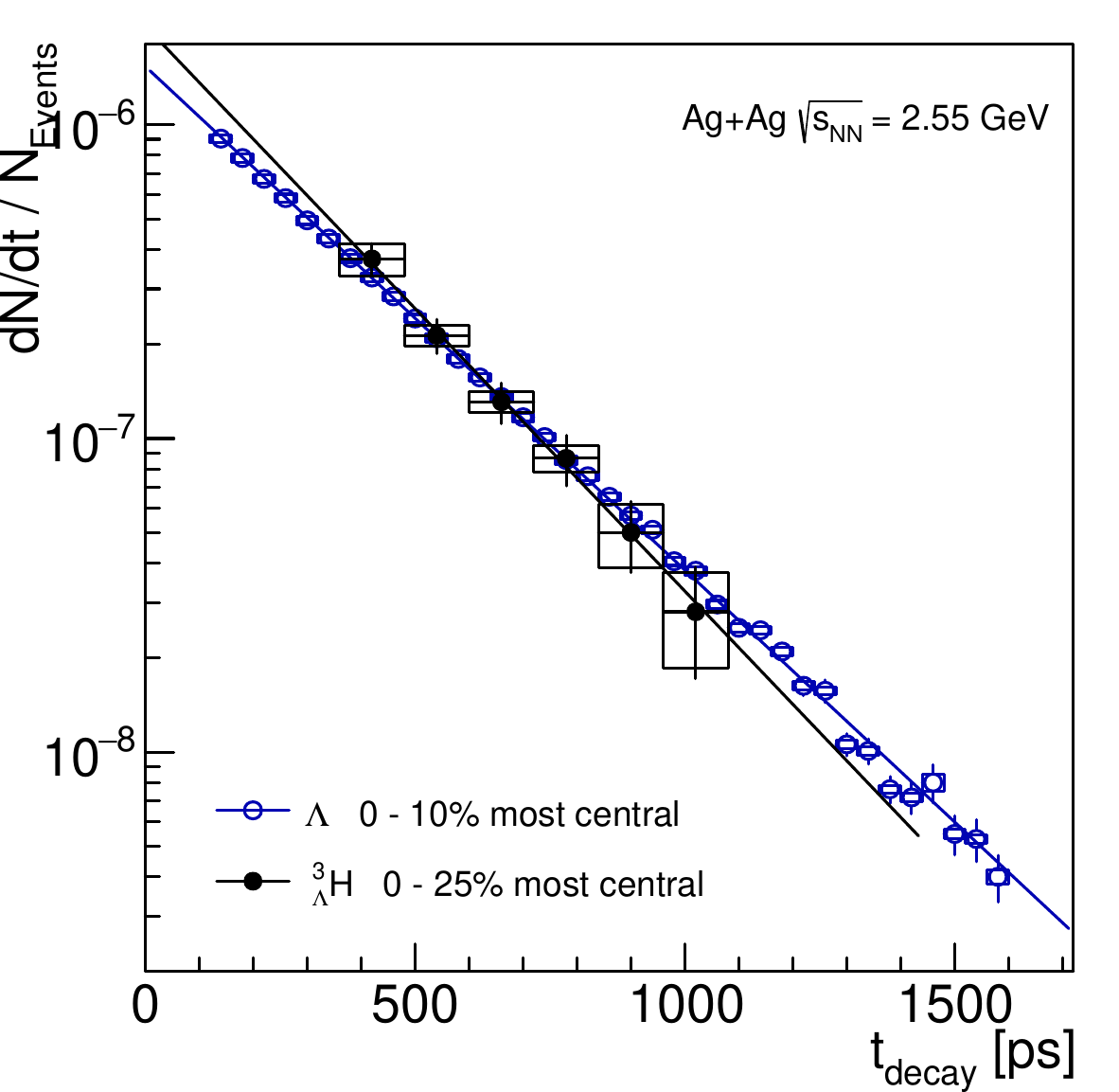}
        \includegraphics[width=\linewidth]{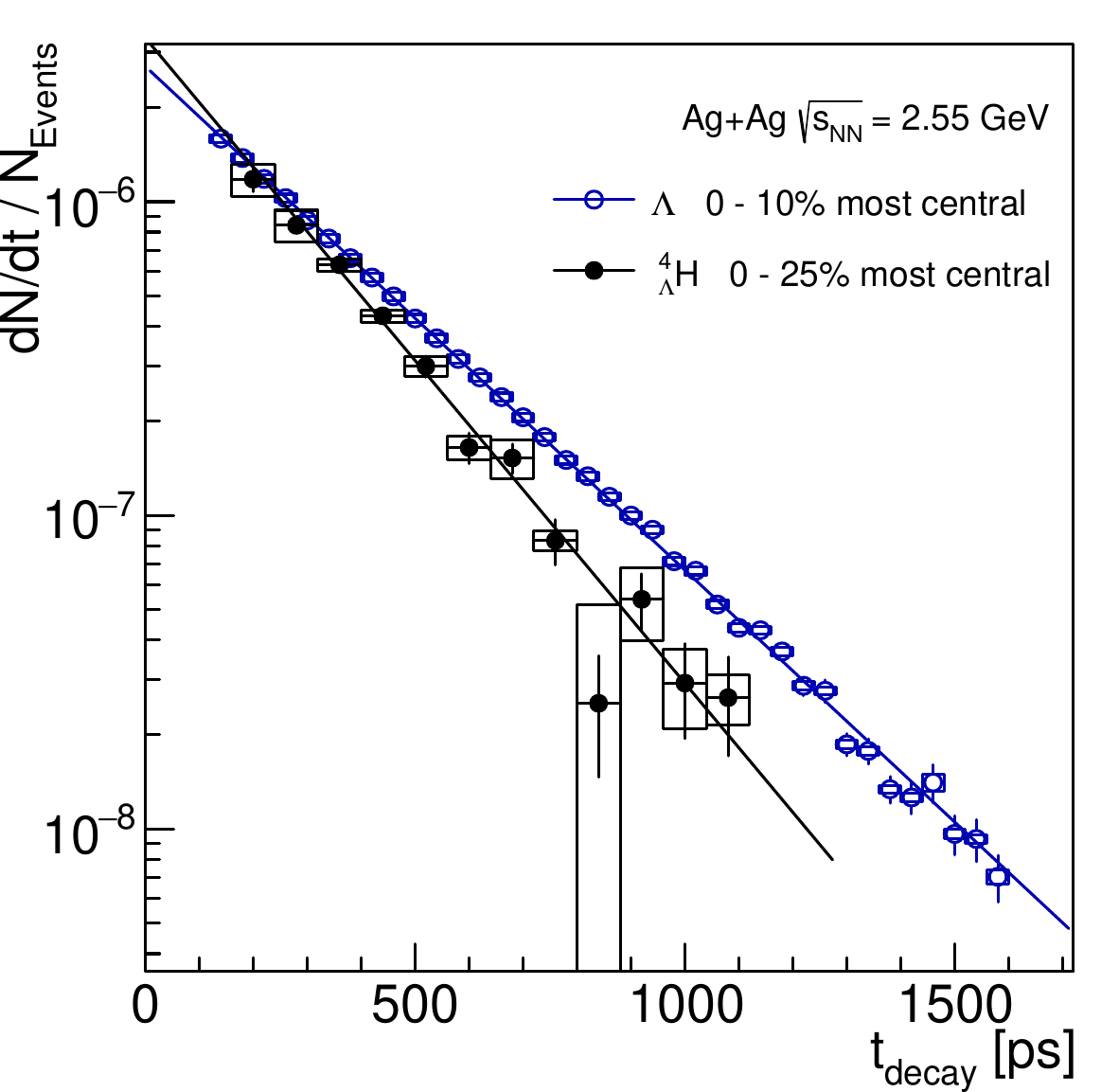}
        \caption{Decay curves of \Hypertriton\ (upper plot, black dots) and \HyperheliumFour\ (lower plot, black dots). For comparison also the decay curve for \LambdaN\ hyperons (blue dots) is plotted. The decay curves are fitted with an exponential (solid curve) to extract values for the mean lifetime $\tau$, see text for details.}
        \label{fig:LifeMeas}
    \end{figure}
    To obtain decay curves for both hypernuclei and to extract their corresponding average lifetime \(\tau\), the decay time \(t\) of each hypernucleus candidate is calculated from its decay length \(l\), velocity \(\beta\), and Lorentz factor \(\gamma\). The \Hypertriton{} is identified as described above in six time intervals of 120\,ps width each, ranging from 360 to 1080\,ps, and the \HyperhydrogenFour{} in twelve intervals of 80\,ps width, ranging from 160 to 1120\,ps.
    
    The extracted signal yields are corrected for acceptance and efficiency losses analogously to the procedure used in the multi-differential analysis in transverse momentum and rapidity. Note that for a given decay time \(t\), the underlying kinematic distributions significantly affect the distribution of decay lengths \(l\), and thus the integrated acceptance and efficiency correction factors. Therefore, realistic kinematic distributions as input to the simulations are essential.

The capability of HADES to measure decay curves and reliably extract lifetimes is validated using \(\Lambda\) hyperons, which exhibit similar decay topologies. The extracted mean lifetime \(\tau_{\Lambda} = (272 \pm 1{\mathrm{(stat)}} \pm 11{\mathrm{(sys)}})\,\mathrm{ps}\) ~\cite{Simon_Hyp}, agrees with the PDG value of $(263 \pm 2)$ ps.

The resulting decay curve \(dN/dt\) for the \Hypertriton{} (black dots) is shown in the upper panel of \superref{Fig.}{fig:LifeMeas}, and that for the \HyperhydrogenFour{} (black dots) in the lower panel. For comparison, the decay curve of \(\Lambda\) hyperons (blue dots), extracted from the 10\,\% most central \AgAg{} events, is displayed in both cases.

While the \Hypertriton{} data points are in fair agreement with those obtained for \(\Lambda\) hyperons, deviations become apparent in the case of the \HyperhydrogenFour{}. This can be quantified by fitting an exponential function (black solid curves in \superref{Fig.}{fig:LifeMeas}) of the following form to the decay distributions of both hypernuclei to extract their respective mean lifetimes:

\begin{equation}
    \frac{dN}{dt} = N_0 \exp\left(-\frac{t}{\tau}\right),
    \label{eq:expfit}
\end{equation}
where \(N_0\) denotes the initial yield at \(t = 0\) and \(\tau\) is the mean lifetime of the respective hypernucleus.

The extracted lifetime for the \Hypertriton{} is \(\tau_{{}^{3}_{\Lambda}\mathrm{H}} = (239 \pm 23{\mathrm{(stat)}} \pm 18{\mathrm{(sys)}})\,\mathrm{ps}\), and for the \HyperhydrogenFour{} \(\tau_{{}^{4}_{\Lambda}\mathrm{H}} = (209 \pm 7{\mathrm{(stat)}} \pm 10{\mathrm{(sys)}})\,\mathrm{ps}\).
The systematic uncertainties were determined by evaluating the changes in the extracted mean lifetime \(\tau\) resulting from the variations of the analysis parameters as described when discussing the \(d^{2}N/(dp_{\mathrm{T}}\,dy)\) distributions.

While the lifetime for the \Hypertriton{} is consistent on the 1$\sigma$ level with that of the free \LambdaN, the \HyperhydrogenFour\ lifetime shows a 4.5 $\sigma$ deviation from the free \LambdaN\ lifetime.

The extracted mean lifetime values \(\tau_{{}^{3}_{\Lambda}\mathrm{H}}\) and \(\tau_{{}^{4}_{\Lambda}\mathrm{H}}\) are compared with the available world data:  
for the \Hypertriton{}~\cite{bib:Hypertriton1964, bib:Hypertriton1968, bib:Hypertriton1969, bib:Hypertriton1970_1, bib:Hypertriton1970_2, bib:Hypertriton1973, bib:STAR_Hypertriton2010, bib:HypHI_Hypertriton2013, bib:ALICE_Hypertriton2016, bib:STAR_Hypertriton2018, bib:ALICE_Hypertriton2019, bib:STAR_Hypertriton2022, ALICE:2022rib} and for the \HyperhydrogenFour{}~\cite{bib:Hyperhydrogen1962, bib:Hypertriton1964, bib:Hyperhydrogen1965, bib:Hypertriton1969, bib:Hyperhydrogen1992, bib:Hyperhydrogen1995, bib:HypHI_Hypertriton2013, bib:STAR_Hypertriton2022, bib:Hyperhydrogen2023}, as shown in \superref{Fig.}{fig:LifeComp}.

In both panels, the blue line corresponds to the lifetime of a free \(\Lambda\) hyperon, while the black line indicates the current world average — shown separately for the \Hypertriton{} (upper panel) and the \HyperhydrogenFour{} (lower panel). The shaded area denotes the corresponding one-\(\sigma\) uncertainty interval around the world average.

In both cases, the value measured by HADES agrees within errors with the world average. The achieved measurement precision is competitive with other recent results and significantly enriches the available world data.

Including the HADES data into the world data average, a deviation of \(3.9\,\sigma\) from the free \(\Lambda\) lifetime is observed for the \Hypertriton{}, and \(6.3\,\sigma\) in the case of the \HyperhydrogenFour{}.

    \begin{figure}
        \centering
        \includegraphics[width=\linewidth]{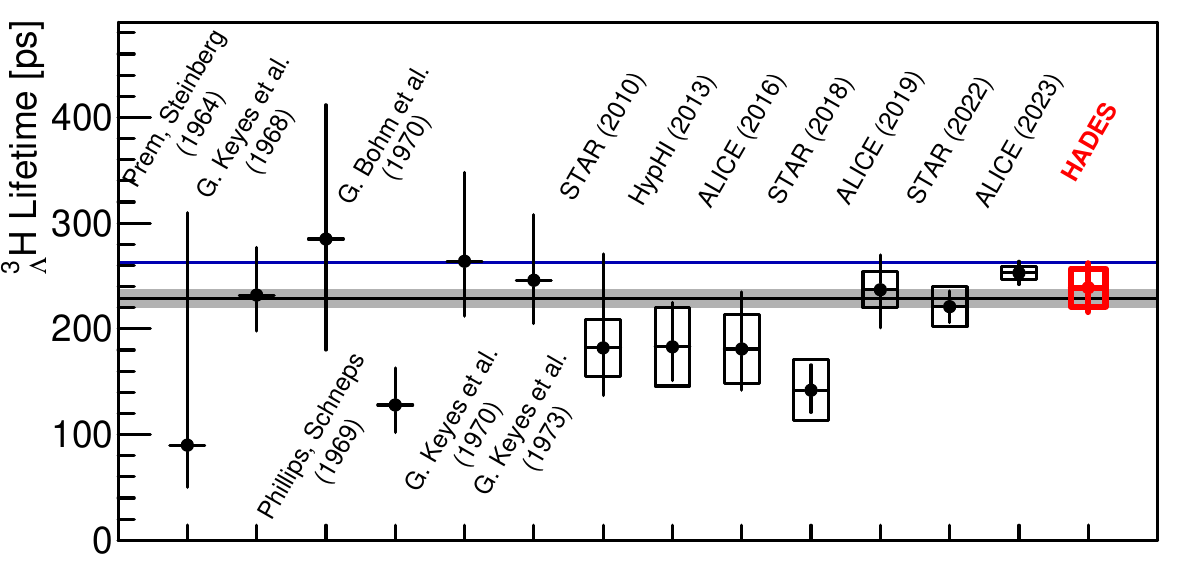}
        \includegraphics[width=\linewidth]{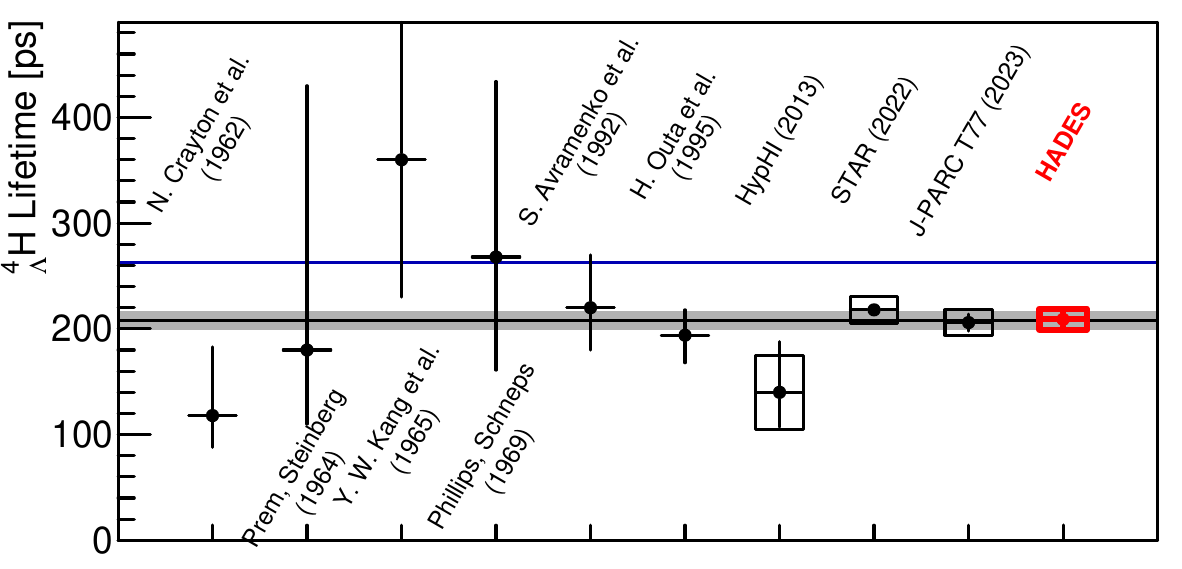}
        \caption{Comparison of the measured lifetimes with available world data. In both cases, the blue horizontal line indicates the lifetime of a free \(\Lambda\) hyperon, while the black line represents the current world average for the respective hypernucleus: \(^{3}_{\Lambda}\mathrm{H}\) (top) and \(^{4}_{\Lambda}\mathrm{H}\) (bottom). The shaded band shows the corresponding one-\(\sigma\) uncertainty interval around the world average.}
        \label{fig:LifeComp}
    \end{figure}
    \clearpage

\section{Summary}\label{sec:summary}

We report the first observation of \Hypertriton{} and \HyperhydrogenFour{} produced in heavy-ion collisions at \(\sqrt{s_{\mathrm{NN}}} = 2.55~\mathrm{GeV}\), reconstructed via their two-body mesonic decay channels in central \AgAg{} collisions. The hypernuclei are identified through their topological decay signatures, with background suppression enhanced by an artificial neural network.

The measured multi-differential yields \(d^2N/(dp_{\mathrm{T}}\,dy)\) and the integrated rapidity distributions \(dN/dy\) exhibit bell-shaped structures centered around mid-rapidity. The hypernuclear yields at mid-rapidity are compared to the existing world data. In particular, our measurement doubles the available world data points for \HyperhydrogenFour{}.
The yield of \HyperhydrogenFour\ is equal to or exceeds that of \Hypertriton, which contrasts the measurement from the STAR collaboration at \sqrtSNN~=~3~GeV which supports a secnario, in which 
excited $A{=}4$ states enhance $\HyperhydrogenFour$ formation at high $\mu_B$.
Decay time distributions are reconstructed and corrected for acceptance and efficiency effects to extract the mean lifetimes. While the extracted lifetime of the \Hypertriton{} is consistent with that of the free \(\Lambda\) hyperon within uncertainties, the \HyperhydrogenFour{} exhibits a statistically significant deviation of approximately \(4.9\,\sigma\). Including the present results, the world data show a deviation from the free \(\Lambda\) lifetime of \(3.9\,\sigma\) for the \Hypertriton{} and \(6.3\,\sigma\) for the \HyperhydrogenFour{}.
\newline

\textbf{Acknowledgment}

SIP JUC Cracow, Cracow (Poland), 2017/26/M/ST2/00600, 2023/49/B/ST2/00652 (OPUS), SONATA-BIS 2023/50/E/ST2/00673; TU Darmstadt, Darmstadt (Germany), VH-NG-823, DFG GRK 2128, DFG CRC-TR 211, BMBF:05P18RDFC1, HFHF (Campus Darmstadt), ELEMENTS 500/10.006, GSI F\&E, EMMI GSI Darmstadt; Goethe-University, Frankfurt (Germany), BMBF:05P12RFGHJ, GSI F\&E, HFHF (Campus Frankfurt), EMMI GSI Darmstadt, ELEMENTS 500/10.006; JLU Giessen, Giessen (Germany), BMBF:05P12RGGHM; IJCLab Orsay, Orsay (France), CNRS/IN2P3; NPI CAS, Rez, Rez (Czech Republic), MSMT LM2023060, MSMT OP JAK CZ.02.01.01/00/23\_015/0008181.

We acknowledge the contribution from the following colleagues:: Alexander Belyaev, Oleg Fateev, Alexander Ierusalimov, Vladimir Ladygin, Sergei Reznikov, Alexander Zinchenko, Dmitriy Borisenko, Marina Golubeva, Fedor Guber, Alexander Ivashkin, Nikolay Karpushkin, Sergey Morozov, Oleg Petukhov, Andrei Reshetin, Arseniy Shabanov, Vladimir Khomyakov, Alexander Lebedev, Alexander Zhilin, Oleg Golosov, Mikhail Mamaev, Arkadiy Taranenko.

\end{document}